\documentclass[sigconf]{acmart}
\AtBeginDocument{%
  }

\copyrightyear{2025}
\acmYear{2025}
\setcopyright{rightsretained}
\acmConference[EICS Companion '25]{Companion Proceedings of the 17th ACM SIGCHI Symposium on Engineering Interactive Computing Systems}{June 23--27, 2025}{Trier, Germany}
\acmBooktitle{Companion Proceedings of the 17th ACM SIGCHI Symposium on Engineering Interactive Computing Systems (EICS Companion '25), June 23--27, 2025, Trier, Germany}\acmDOI{10.1145/3731406.3731973}
\acmISBN{979-8-4007-1866-3/2025/06}

\usepackage{subcaption}

\begin{document}

\title{Evaluating Joint Attention for Mixed-Presence Collaboration on Wall-Sized Displays}

\author{Adrien Coppens}
\email{adrien.coppens@list.lu}
\orcid{0000-0002-2841-6708}
\affiliation{%
  \institution{Luxembourg Institute of Science and Technology}
  \city{Esch-sur-Alzette}
  \country{Luxembourg}
}

\author{Valérie Maquil}
\email{valerie.maquil@list.lu}
\orcid{0000-0002-0198-3729}
\affiliation{%
  \institution{Luxembourg Institute of Science and Technology}
  \city{Esch-sur-Alzette}
  \country{Luxembourg}
}


\begin{abstract}
To understand and quantify the quality of mixed-presence collaboration around wall-sized displays, robust evaluation methodologies are needed, that are adapted for a room-sized experience and are not perceived as obtrusive. In this paper, we propose our approach for measuring joint attention based on head gaze data.
We describe how it has been implemented for a user study on mixed presence collaboration with two wall-sized displays and report on the insights we gained so far from its implementation.
\end{abstract}





\begin{teaserfigure}
\centering
    \includegraphics[width=.9\textwidth]{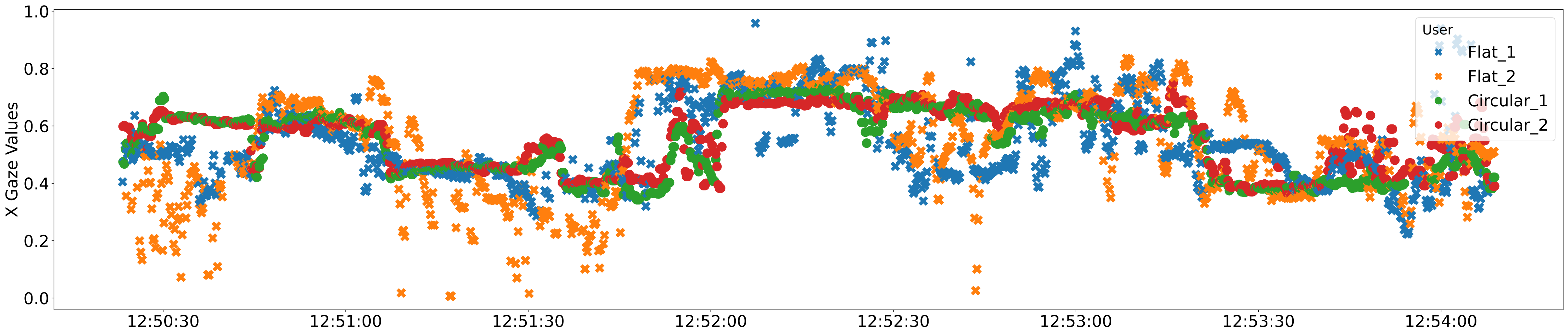}
    \caption{Scatter plot of horizontal gaze values for all four participants over time, for the mixed-presence part of a session.}
    \label{fig:scatterplot}
    \Description{The scatter plot shows the evolution of horizontal gaze values for the four participants over time. Although there are some discrepancies, especially for one of the participants that seem to intermittently produce low values than all three others (therefore looking to the left of the display), it seems like the values are correlated i.e. they roughly follow the same evolution. We can also see that the participants from the circular display seem to display even higher correlation between them than with the other two participants from the flat display. It also seems in general that the values for the participants from the flat display vary more (more variance).}
\end{teaserfigure}

\maketitle

\section{Introduction} 

Given their large size accommodating multiple users, Wall-Sized Displays (WSDs) were shown to naturally afford room-scale interactions as well as collocated collaboration~\cite{ni2006survey}. 
However, remote and mixed-presence collaboration (with a mixture of collocated and remote participants) through WSDs remains under-explored, and poses additional challenges as efficient collaboration requires a good understanding of each other's interaction with the shared workspace (i.e. Workspace Awareness~\cite{gutwin2002descriptive}). One important aspect of that is \textit{Gaze} (where users are looking), which has recently been categorised as \textit{Attention} in the specific context of remote collaboration through WSDs~\cite{coppens2024workspace}. A common measure derived from it is joint attention (sometimes called joint visual attention or joint focus of attention) as it is considered as a collaboration indicator~\cite{schneider2018leveraging}.

To measure collaborative processes such as gaze and joint attention, robust evaluation methodologies are needed~\cite{scott2003system, mateescu2021collaboration}.
%
%
%
Gaze-related research largely relies on eye tracking sensors~\cite{novak2024eye} which are usually considered reliable to evaluate where someone is looking and typically come into two form factors: as bars that one places at a fixed position, or as glasses that users can wear.
That being said, eye tracking bars are not suited for room-scale interactions (such as those around WSDs) as users need to be close to these bars for them to work. In contrast, head-mounted sensors are more suited for such environments, and have indeed been used to analyse the behaviour of users around large vertical displays in both collocated and remote collaboration settings~\cite{wisiecka2023supporting}. However, having to wear such devices might feel obtrusive to users, in particular in mixed-presence settings, and this may influence the behaviour of users. 
Although less accurate to analyse gaze, an alternative option is to track head position instead, as this can be achieved with ``extrinsic'' depth cameras, such as the Kinect family of sensors, and still provides a good idea of where a person's attention is~\cite{stiefelhagen1999gaze}.

To date, most of the research efforts have been spent on measuring joint attention for pairs of users~\cite{jermann2012duet}. In this poster, we focus on an collaborative setting with in total four users and describe our approach for detecting and quantifying joint attention using Azure Kinect and the \textbackslash psi framework. The plots we show focus on one particular session used as a running example.

\begin{figure*}[t]
    \centering
    \includegraphics[width=.34\linewidth]{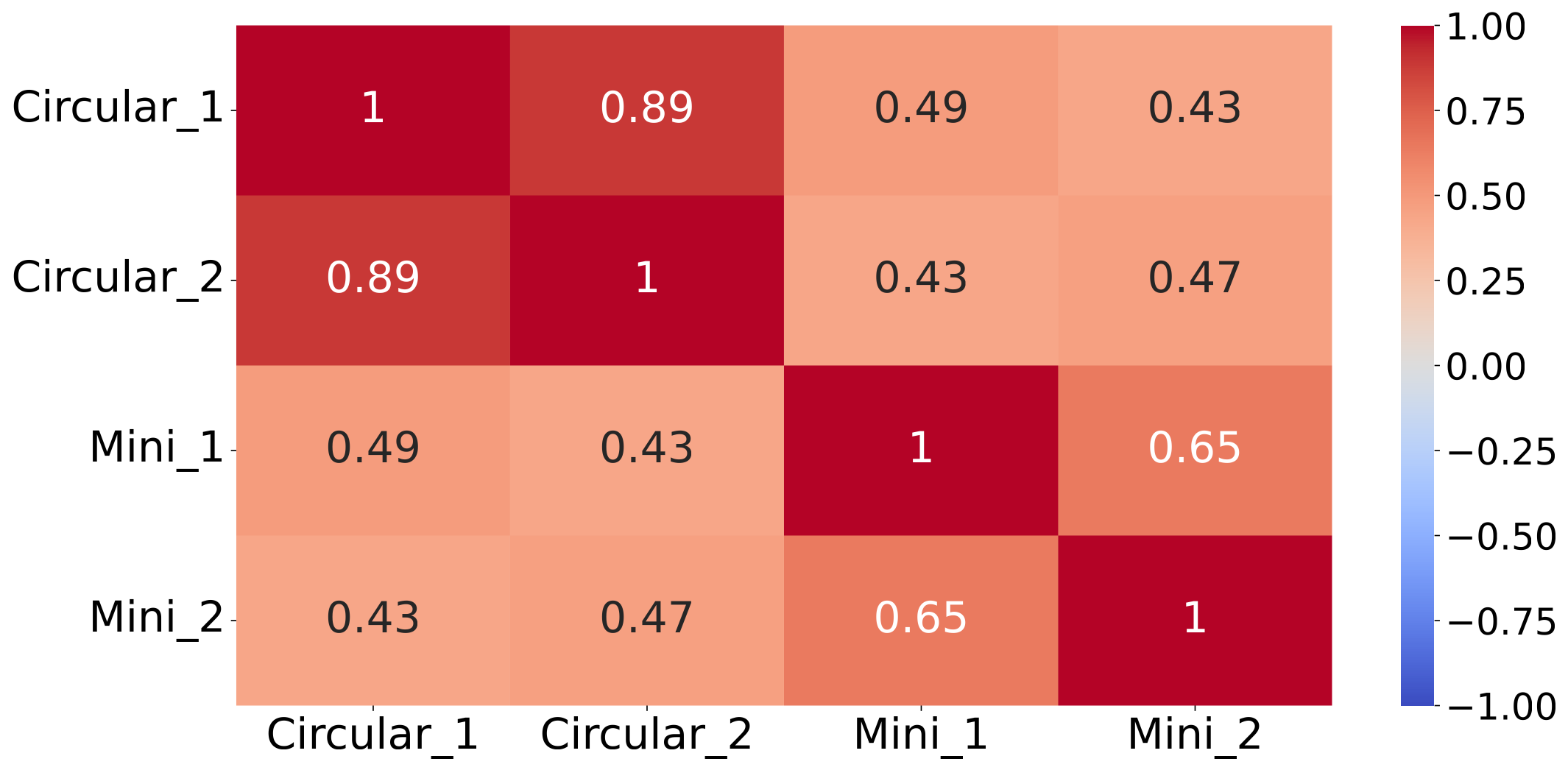}
    \Description{The correlation matrix shows pairwise spearman coefficients, showing values for all possibles pairs of participants. The legend on the side of the figure clarifies that positive values are red while negative ones are blue, and that lighter colours are close to 0, while darker colours are close to 1 (or -1 for the negative part). So dark red means high positive correlation, light red means low positive correlation, light blue means low negative correlation, and dark blue means high negative correlation. All cells of the matrix are coloured in varying shades of red. Obviously the diagonal (which corresponds to the correlation between each user and her/himself) has a value of 1 and the darker possible shade of red. We visually notice that colours are darker (and therefore correlation values are higher) for pairs of participants from the same side, using the same display. The values are 0.89 and 0.65 respectively for participants from the circular display and the flat display, while values for pairs of participants from different sides are all between 0.43 and 0.49.}
    \hspace{0.3em}
    \includegraphics[width=.555\linewidth]{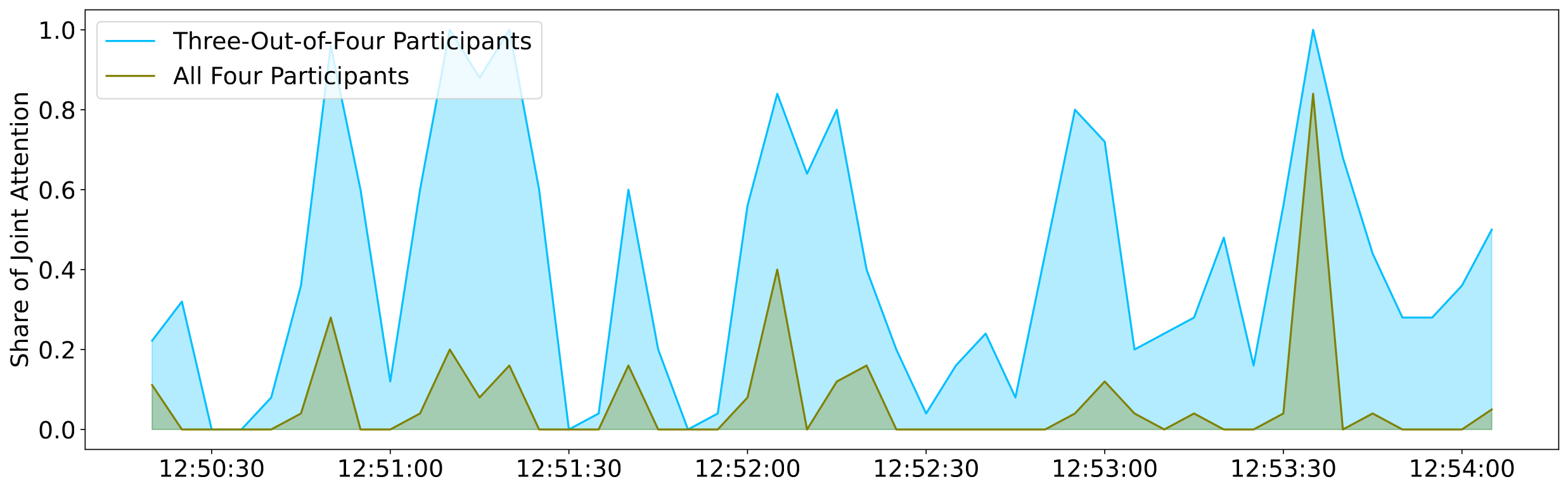}
    \Description{The plot shows the evolution of the share of joint attention over time, using either "three-out-of-four" or "all-four" participants. We see that the values for "three-out-of-four" are always higher, as expected. In fact, the values for "all-four" participants are in general very low, with only one high spike towards the end of the session. Regarding "three-out-of-four" values, there regularly are high spikes but also periods with very low values (some in between but in general many variations and extreme values).}
    \caption{Correlation matrix for horizontal gaze values (left) and share of joint attention over time (right) for the chosen session.}
    \label{a-two-column-figure}
\end{figure*}

\section{Gathering and Normalising Data}

Our focus was on mixed-presence experiments, with two pairs of users working on a task from two separate rooms. 
In order to record such experiments, we relied on Azure Kinect sensors and the \textbackslash psi framework~\cite{bohus2021platform}; in particular a subset of a body tracking and awareness transmission pipeline~\cite{coppens2024supporting}. This allowed us to gather head gaze data in both rooms thanks to two computers.
We later on reprocessed the recordings to produce tabular data (in CSV format) we could feed to Python scripts to compute metrics and draw plots. 


As we had two files for each session (one per room), we merged them into a single one per session. However, we had to go through a few steps to normalise the data, which we describe here.
First, we applied a time offset to make up for the time difference (the offset was determined based on on-screen events and audio that could be observed or heard from both sides).  
We then trimmed the data, only keeping rows within the mixed-presence part of the experiment. 

As the identification information from the body tracking (skeleton id) was sometimes inconsistent (e.g. a user coming back within tracking range getting a new id) and because other persons (e.g. the facilitator running the experiment) were sometimes within the tracking range, we manually merged 
and filtered the data for each session. 
As tracking was sometimes lost (e.g. for users going out of range), we also used linear interpolation to deal with the missing gaze data to make sure we had values for all users at all times.

Finally, as the displays on both sides had different resolutions, we normalised gaze values between 0 and 1 for both axes (so a gaze target at $(0,0)$ corresponds to the lower left corner of the display). 

\section{Evaluating Data}

Based on the time series of gaze values (target location on the display) for each user and each session, we started our joint attention analysis by creating scatter plots for both axes. As illustrated in Figure~\ref{fig:scatterplot} showing the evolution of horizontal coordinate values for the given session, this allowed us to get a grasp on the data at hand.

As there seem to be some correlation between data for different users at least for this session, we continued our exploratory analysis by computing correlation matrices for x values between all pairs of participants. We opted to use Spearman's $\rho$ coefficient (based on the rankings of the values) rather than Pearson's, as it is more robust to outliers~\cite{de2016comparing} and better captures non-linear relationships.

As shown in the left part of Figure~\ref{a-two-column-figure}
, the values we obtained were sufficient to observe some interesting results for that session at least. It seems that horizontal gaze values are indeed correlated, in particular between participants from the same room. This essentially means that collocated collaborators tend to look at the same screen. Even though the $\rho$ values are lower and the correlation is therefore less pronounced for remote collaborators, this is also the case for cross-site values. Since that particular session was using a screen-wide attention cue (showing which screen both the local and remote participants were looking at) and because head movements from local collaborators lead to natural awareness cues which are known to be effective, we were indeed expecting such results.


Aim of the next step was to actually quantify joint attention. For participants to be considered as demonstrating joint attention, their gaze targets need to be sufficiently close to each other. We therefore used the Euclidean distance as metric to assess joint attention. 
For the experiment we gathered the data from, participants were expected to look at art pieces. In that context, it is natural to define joint attention as "looking at the same art piece", so we considered the maximum dimensions an art piece can have to set the distance threshold under which we consider targets to correspond to joint attention.
Taking into account the web-based layout of our system, which is used with two different configurations of WSDs, we used a normalized value, based on percentages of the displays' size.


As we are interested in mixed-presence joint attention, we calculate for each timestamp the maximum distance among the values for all pairs of participants. If that maximum distance is below the threshold, then we consider that there is joint attention as all four participants' gaze targets are close, within an area that could correspond to an art piece.
In addition, we computed the maximum distance between the best triplet of participants (removing the participant who leads to the worst overall value).
Based on both of these types of values (all-four and three-out-of-four maximum distances), we computed the the share of time spent below threshold over 5s time windows. This gives us a view on the evolution of joint attention over time, as shown in the right part of Figure~\ref{a-two-column-figure}.


\section{Discussion}
These early results look promising as a basis for gathering insights from the data we collected. To go further, we will need to process other sessions and aggregate the data from all sessions, looking for patterns, mainly in sessions using the same experimental condition.

Rather than using the entire mixed-presence part of each session, it could be interesting to trim the data even further and focus on time windows around deictic references or descriptive explanations.

Our dataset is available online~\cite{dataset}.



\begin{acks}
This research was funded by FNR under ReSurf (Grant C21/IS/15883550).

\end{acks}

\bibliographystyle{ACM-Reference-Format}

\end{document}